\documentclass{PoS}

\title{\code{}  library for rare decays in the MSSM}
\usepackage{amsmath}
\usepackage{amssymb}
\def\code{{\tt SUSY\_FLAVOR}}
\ShortTitle{\code{}}

\author{\speaker{Andreas Crivellin} \thanks{This work is supported by
    the Swiss National Science Foundation (SNF).}\\
    Albert Einstein Center for Fundamental Physics, \\
    Institute for Theoretical Physics, University of Bern.\\ 
    E-mail: \email{crivellin@itp.unibe.ch}}

\author{Janusz Rosiek\thanks{This work is supported in part by the EU
    the contract PITN-GA-2009-237920 UNILHC and by the National
    Science Centre in Poland under the research grants DEC-
    2011/01/M/ST2/02466 and DEC-2012/05/B/ST2/02597.}\\
    Institute of Theoretical Physics\\ 
    Physics Department, University of Warsaw\\ 
    E-mail: \email{janusz.rosiek@fuw.edu.pl}}

\abstract{\code{} is a FORTRAN code calculating over 30 low-energy
  flavour- and CP-related observables in the $R$-parity conserving
  MSSM. The code admits for the most general flavour structure of the
  SUSY breaking terms and complex flavour-diagonal couplings.  It
  includes the numerically important resummation of chirally enhanced
  effects and it is fast enough for scanning over a large
  SUSY-parameter space.  The program can be obtained from {\tt
    http://www.fuw.edu.pl/susy\_flavor}.}

\FullConference{The European Physical Society Conference on High
  Energy Physics -EPS-HEP2013\\ 18-24 July 2013\\ Stockholm, Sweden}

\begin{document}

\code{} v2~\cite{Rosiek:2010ug} is a fast FORTRAN 77 program which
calculates over 30 low-energy flavour- and CP-related observables in
the MSSM taking the fully general set of SUSY parameters as an
input. The program works in the following steps:
\begin{enumerate}
\item Parameter initialization: the user sets the SUSY breaking terms
  and other MSSM parameters using the SLHA2 format. SM parameters and
  hadronic matrix elements can also be modified.
\item Physical masses and the mixing angles: the physical spectrum of
  the MSSM is calculated by exact numerical diagonalization of the
  relevant mass matrices.
\item Resummation of the chirally enhanced effects: \code{} resums all
  chirally enhanced effects to the relation between the fermion masses
  and the MSSM Yukawa couplings and calculates the effective
  Higgs-fermion-fermion and fermion-sfermion-gaugino
  vertices~\cite{Crivellin:2010er}.
\item Wilson coefficients at the SUSY scale: \code{} calculates the
  virtual SUSY corrections to the Wilson coefficients of the
  effective Hamiltonian used to compute the flavour observables.
\item Strong corrections: the matrix elements of the effective
  Hamiltonian are calculated using the data from lattice QCD. The user
  can update QCD input if necessary.
\item Evaluation of the physical observables: Table~\ref{tab:proc}
  give the list of observables calculated by \code.
\end{enumerate}

The resummation of the chirally enhanced corrections, including the
threshold corrections to Yukawa couplings and CKM matrix elements, is
an important new feature added to \code{} in version $2.0$. Such
corrections arise in the case of large values of $\tan\beta$ or large
trilinear SUSY-breaking terms. They formally go beyond the $1$-loop
approximation, but should be included due to their numerical
importance. Implementation of the resummation in \code{} follows the
systematic approach of Refs.~\cite{Crivellin:2010er} and takes into
accounts all contributions involving sfermions, gluino, chargino,
neutralino and Higgs boson exchanges.

\code{} v2 is an universal and easy to use numerical tool which can be
used by both experimentalists and theoreticians working in the field
of supersymmetric flavour physics. The program can be obtained from
{\tt http://www.fuw.edu.pl/susy\_flavor}.

\begin{table}[htbp]
\renewcommand{\arraystretch}{1.2}
\begin{center}
\begin{tabular}{|lcr|}
\hline
Observable & Most stringent constraints on &Experiment \\ \hline
\multicolumn{3}{|l|}{$\Delta F=0$} \\ \hline

$\frac{1}{2}(g-2)_e$ & $\rm{Re}\left[\delta^{\ell\,LR,RL}_{11}\right]$
&$(1 159 652 188.4 \pm4.3) \times 10^{-12}$ \\

$\frac{1}{2}(g-2)_\mu$ &
$\rm{Re}\left[\delta^{\ell\,LR,RL}_{22}\right]$
&$(11659208.7\pm8.7)\times10^{-10}$ \\

$\frac{1}{2}(g-2)_\tau$ &
$\rm{Re}\left[\delta^{\ell\,LR,RL}_{33}\right]$ &$<1.1\times 10^{-3}$
\\

$|d_{e}|$(ecm) & $\rm{Im}\left[\delta^{\ell\,LR,RL}_{11}\right]$
&$<1.6 \times 10^{-27}$ \\

$|d_{\mu}|$(ecm) & $\rm{Im}\left[\delta^{\ell\,LR,RL}_{22}\right]$
&$<2.8\times 10^{-19}$ \\

$|d_{\tau}|$(ecm) & $\rm{Im}\left[\delta^{\ell\,LR,RL}_{33}\right]$
&$<1.1\times 10^{-17}$ \\

$|d_{n}|$(ecm) & $\rm{Im}\left[\delta^{d\,LR,RL}_{11}\right]$,
$\rm{Im}\left[\delta^{u\,LR,RL}_{11}\right]$ &$<2.9 \times
10^{-26}$\\ \hline

\multicolumn{3}{|l|}{$\Delta F=1$}\\ \hline 

$\mathrm{Br}(\mu\to e \gamma)$ & $\delta^{\ell\,LR,RL}_{12,21}$,
$\delta^{\ell\,LL,RR}_{12}$ & $<2.4 \times 10^{-12}$\\

$\mathrm{Br}(\tau\to e \gamma)$ & $\delta^{\ell\,LR,RL}_{13,31}$,
$\delta^{\ell\,LL,RR}_{13}$ & $<3.3\times 10^{-8}$\\

$\mathrm{Br}(\tau\to \mu \gamma)$ & $\delta^{\ell\,LR,RL}_{23,32}$,
$\delta^{\ell\,LL,RR}_{23}$ & $<4.4\times 10^{-8}$\\

$\mathrm{Br}(K_{L }\to \pi^{0} \nu \nu)$ & $\delta^{u\,LR}_{23},
\delta^{u\,LR}_{13}\times\delta^{u\,LR}_{23}$ & $< 6.7\times10^{-8}$
\\

$\mathrm{Br}(K^{+}\to \pi^{+} \nu \nu)$ & sensitive to
$\delta^{u\,LR}_{13}\times\delta^{u\,LR}_{23}$ &
$17.3^{+11.5}_{-10.5}\times 10^{-11}$ \\

$\mathrm{Br}(B_{d}\to e e)$ & $\delta^{d\,LL,RR}_{13}$ & $<1.13\times
10^{-7}$\\

$\mathrm{Br}(B_{d}\to \mu \mu)$ & $\delta^{d\,LL,RR}_{13}$ &
$<8\times 10^{-10}$\\

$\mathrm{Br}(B_{d}\to \tau \tau)$ & $\delta^{d\,LL,RR}_{13}$ &
$<4.1\times10^{-3}$ \\

$\mathrm{Br}(B_{s}\to e e)$ & $\delta^{d\,LL,RR}_{23}$ & $<7.0\times
10^{-5}$\\

$\mathrm{Br}(B_{s}\to \mu \mu)$ & $\delta^{d\,LL,RR}_{23}$ &
$3.2^{+1.5}_{-1.2}\times 10^{-9}$\\

$\mathrm{Br}(B_{s}\to \tau \tau)$ & $\delta^{d\,LL,RR}_{23}$ & $--$\\

$\mathrm{Br}(B_{s}\to \mu e)$ &
$\delta^{d\,LL,RR}_{23}\times\delta^{\ell\,LL,RR}_{12}$ & $<2.0\times
10^{-7}$\\

$\mathrm{Br}(B_{s}\to \tau e )$ &
$\delta^{d\,LL,RR}_{23}\times\delta^{\ell\,LL,RR}_{13}$ & $<2.8\times
10^{-5}$\\

$\mathrm{Br}(B_{s}\to \mu \tau)$ &
$\delta^{d\,LL,RR}_{23}\times\delta^{\ell\,LL,RR}_{23}$ & $<2.2\times
10^{-5}$\\

$\mathrm{Br}(B^+\to \tau^+ \nu)$ & -- & $(1.65\pm 0.34)\times10^{-4}$
\\

$\mathrm{Br}(B_{d}\to D\tau \nu)/\mathrm{Br}(B_{d}\to Dl \nu)$ & -- &
($0.407 \pm 0.12 \pm 0.049)$ \\

$\mathrm{Br}(B\to X_{s} \gamma)$ & $\delta^{d\,LL,RR}_{23}$ for large
$\tan\beta$, $\delta^{d\,LR}_{23,32}$ & $(3.52\pm 0.25) \times
10^{-4}$\\

\hline

\multicolumn{3}{|l|}{$\Delta F=2$}\\ \hline 

$|\epsilon_{K}|$ & $\rm{Im}\left[(\delta^{d\,LL,RR}_{12})^2\right]$,
$\rm{Im}\left[(\delta^{d\,LR}_{12,21})^2\right]$ & $(2.229 \pm
0.010)\times 10^{-3}$ \\

$\Delta M_{K}$ & $\delta^{d\,LL,RR}_{12}$, $\delta^{d\,LR}_{12,21}$ &
$(5.292 \pm 0.009)\times10^{-3}~\mathrm{ps}^{-1}$\\

$\Delta M_{D}$ & $\delta^{u\,LL,RR}_{12}$, $\delta^{u\,LR}_{12,21}$ &
$(2.37^{+0.66}_{-0.71}) \times10^{-2}~\mathrm{ps}^{-1}$\\

$\Delta M_{B_{d}}$ & $\delta^{d\,LL,RR}_{13}$,
$\delta^{d\,LR}_{13,31}$ & $(0.507 \pm0.005)~\mathrm{ps}^{-1}$\\

$\Delta M_{B_{s}}$ & $\delta^{d\,LL,RR}_{23}$,
$\delta^{d\,LR}_{23,32}$ & $(17.77 \pm0.12)~\mathrm{ps}^{-1}$\\ \hline
\end{tabular}

\end{center}
\caption{List of observables calculated by \code{} v2, their currently
  measured values or bounds and the elements of the sfermion mass
  matrices most stringently constrained by the corresponding
  process. \label{tab:proc}}
\end{table}

\end{document}